\documentclass[aps,prb,preprint,groupedaddress]{revtex4}

\begin{document}


\title{Clarifying Einstein's First Derivation for Mass-Energy Equivalence and Consequently Making Ives's Criticism a Void}



\author{R. V. R. Pandya}
\thanks{Corresponding author}
\email[]{vpandya@me.uprm.edu}
\affiliation{Department of Mechanical Engineering, University of Puerto
Rico at Mayaguez, Mayaguez,
PR 00680, USA }
\date{\today}


\begin{abstract}
We study physical situation considered by Einstein (Ann. Physik,
17, 1905) for his first derivation of mass-energy equivalence. Einstein introduced a constant $C$ in his derivation and reasoning surrounding $C$ and equations containing $C$ caused criticism by Ives. Here we clarify Einstein's derivation and obtain a value for constant $C$. The obtained zero value for $C$ suggests alternative explanation for Einstein's derivation and makes Ives's criticism a void and for which details are also presented in this paper.    
\end{abstract}

\maketitle

\section{Introduction}
Different forms of mass-energy equivalence relation existed even
before Einstein's first derivation of the relation
\cite{Einstein05} and which have been reviewed along with other
developments on the relation after the year 1905 (see Ref.
\cite{Fadner88} and references cited therein). The focus here is
on a century old Einstein's first derivation which has remained
persistently debatable for its correctness and completeness
\cite{Fadner88,ST82} after the emergence of Ives' work
\cite{Ives52} suggesting circular argument in the derivation. Here
we show that Einstein's derivation contains hidden but valid condition. Under the presence of the condition we further obtain a value of constant $C$ which Einstein invoked in his derivation (see Eqs. (\ref{ob1}) and (\ref{ob2}) below) leading to criticism by Ives. The obtained zero value for $C$ in the present work makes Ives's criticism a void and is also shown in this paper. We first describe Einstein's
derivation briefly along with our important notes written in \textit{italics}, then present the analysis on hidden
condition, value of $C$ and analysis making Ives's criticism a void.

\section{Einstein's Derivation} Consider a `stationary' reference frame $S_s$ with coordinate axes $(x,y,z)$ and another reference
frame $S_v$ with axes ($\xi,\eta,\zeta$) having constant translational velocity ${\bf v}$ as measured in $S_s$. Also, consider coordinate axes of $S_v$ to be parallel to coordinate axes of $S_s$ and origin of $S_v$ in translational motion along the $x$ axis of $S_s$ with velocity magnitude $|{\bf v}|=v$.
Consider a body of mass $M_s$ at rest in $S_s$ at some elevation and at some instance it emits in two opposite
directions (along the $x$ axis) equal quantity of light having energy $L/2$ where $M_s$ and $L$ are measured in $S_s$.
The conservation of energy principle for this situation in $S_s$ can be written as
\begin{equation}
E_0=E_1+\frac{L}{2}+\frac{L}{2}
\label{e1}
\end{equation}
where $E_0$ and $E_1$ are, respectively, total energy of the body
before and after the emission of the light as measured in $S_s$. 

\textit{It should be noted that Einstein did not consider any gravitational field in his derivation, otherwise gravitation effect on $L$ should be included in Eq. (\ref{e1}). Which means that $L$ should be replaced by 
$L(1+\frac{\Phi}{c^2})$ where $\Phi$ is gravitational potential at the location of the body and $c$ is speed of light \cite{Einstein11}. So Einstein derivation of mass-energy equivalence is valid in the absence of gravitational field. In fact, Einstein derived gravitation of energy afterward in the year 1911. The derivation of mass-energy equivalence relation with inclusion of effect of gravitational field will be presented elsewhere.} 

The conservation of energy principle for the body as observed from $S_v$ can be written as
\begin{equation}
H_0=H_1+\frac{L}{\sqrt{1-v^2/c^2}}
\label{e2}
\end{equation}
where $H_0$ and $H_1$ are, respectively, total energy of the body
before and after the emission of the light as measured in $S_v$.
Subtracting Eq.
(\ref{e1}) from Eq. (\ref{e2}) yields
\begin{equation}
(H_0-E_0)-(H_1-E_1)=L[\frac{1}{\sqrt{1-v^2/c^2}}-1]. \label{eq2a}
\end{equation}

Einstein then provided following argument (hereafter, this argument is referred to as $EA$).

$EA$: {``Thus it is clear the difference $H-E$ can differ from the kinetic energy $K$ of the body, with respect to the other system $(\xi,\eta,\zeta)$, only by an additive constant $C$, which depends on the choice of the arbitrary additive constants of the energies $H$ and $E$''}. And he wrote  
\begin{equation}
H_0-E_0=K_0+C, \label{ob1}
\end{equation}
\begin{equation}
H_1-E_1=K_1+C. \label{ob2}
\end{equation}
Here $K_0$ and $K_1$ are, respectively,
kinetic energy of the body before and after the emission of the
light as measured in $S_v$.

\textit{It should be noted that these two Eqs. (\ref{ob1}) and (\ref{ob2}) with constant $C$ written by Einstein have caused much confusion among researchers.  The argument EA of Einstein and Eqs. (\ref{ob1}) and (\ref{ob2}) became cause for Planck's objection and criticism by Ives} \cite{Ives52} 
\textit{suggesting flaw 
(as described in footnote \footnote{Ives's criticism \cite{Ives52}: Using the exact expression for kinetic energies $K_0$ and $K_1$, Ives showed that
$$(H_0-E_0)-(H_1-E_1)=\frac{L}{m_sc^2}(K_0-K_1)$$
and further he considered it as the difference of two relations (similar to Einstein's argument EA), written as
$$H_0-E_0=\frac{L}{m_sc^2}(K_0+C),$$
$$H_1-E_1=\frac{L}{m_sc^2}(K_1+C).$$
Then he wrote ``these are \textit{not} 
$$H_0-E_0=K_0+C,$$
$$H_1-E_1=K_1+C.$$ 
They differ by the multiplying factor $\frac{L}{m_sc^2}$. What Einstein did by setting down these equations (as ``clear'') was to introduce the relation $\frac{L}{m_sc^2}=1$. Now this is the very relation the derivation was supposed to yield.''
})
in the Einstein's derivation. So if we do not invoke the argument and Eq. (\ref{ob1}) and (\ref{ob2}), Ives's criticism becomes void and this related analysis is presented in the next section.}  

Using Eqs. (\ref{ob1}) and (\ref{ob2}), Einstein obtained from Eq. (\ref{eq2a})
\begin{equation}
K_0-K_1=L[\frac{1}{\sqrt{1-v^2/c^2}}-1]
\label{e3}
\end{equation}
Einstein then neglected fourth and
higher orders terms in $v$ in the expansion of right hand side of
Eq. (\ref{e3}) and simplified Eq. (\ref{e3}) to
\begin{equation}
K_0-K_1=\frac{1}{2}\frac{L}{c^2}v^2 \label{e4}
\end{equation}
and concluded ``{{If a body gives off the energy $L$ in the form of radiation,
its mass diminishes by $L/c^2$}}''. We should mention that Stachel and Torretti \cite{ST82}
showed that the approximation involved in Eq. (\ref{e4}) is not required to arrive at the
conclusion when exact expressions for kinetic energies $K_0$ and $K_1$ are used in Eq. (\ref{e3}).

\section{Hidden Condition, Value of $C$ and Voiding Ives's Criticism} The correctness of Einstein's derivation depend on
the correctness of Eqs. (\ref{ob1}) and (\ref{ob2}). Now we obtain hidden condition in Einstein's
derivation under the assumption that Einstein's Eqs. (\ref{ob1}) and (\ref{ob2}) are correct. So if we find the obtained condition to be valid, that would suggest correctness of Eqs. (\ref{ob1}) and (\ref{ob2}) and Einstein's derivation. Then we show that Eq. (\ref{e3}) can be derived without using $EA$ and Eqs. (\ref{ob1}) and (\ref{ob2}) thus voiding Ives's criticism  which is based on $EA$ and Eqs. (\ref{ob1}) and (\ref{ob2}).

Consider the body as a system in thermodynamic sense. 
In general the total energy
of the body (system) is summation of gravitation potential energy $P$, kinetic
energy $K$ and internal energy. In the case of Einstein's derivation, as gravitation potential was not present as pointed out above, we write total energies ($E_0,
E_1, H_0$ and $H_1$) before and after the emission in reference
frames $S_s$ and $S_v$ in terms of internal and kinetic
energies only.

In $S_s$, kinetic energy of the body is zero thus 
total energy of the body before and after the emission can be written as
\begin{eqnarray}
E_0= M_sI_s, \label{vik1}\\
E_1= (M_s-m_s)I_s'.\label{vik2}
\end{eqnarray}
Here $M_s$ is mass of the stationary body before the emission,
$m_s$ is decrease in mass of the body due to the emission, $I_s$
and $I_s'$ are internal energy per unit mass of the body before and after
the emission, respectively, and all are measured in $S_s$.

As measured in $S_v$, we denote the mass of the moving body before
the emission by $M_v$,  decrease in mass of the body due to the emission by $m_v$, 
internal energy per unit mass of the body before and after the emission
by $I_v$ and $I_v'$, respectively. With these notations we can write total
energy of the body before and after the emission as measured in
$S_v$ as
\begin{eqnarray}
H_0=K_0+M_vI_v,\label{pr1}\\
H_1=K_1+(M_v-m_v)I_v'.\label{pr2}
\end{eqnarray}
Subtracting Eq. (\ref{vik1}) from Eq. (\ref{pr1}) and Eq. (\ref{vik2}) from Eq. (\ref{pr2}), we obtain 
\begin{equation}
H_0-E_0=K_0+\Bigl[M_vI_v-M_sI_s\Bigr],\label{pr1ac}
\end{equation}
and
\begin{equation}
H_1-E_1=K_1+\Bigl[(M_v-m_v)I_v'-(M_s-m_s)I_s'\Bigr].\label{pr2ac}
\end{equation}

Now, if Einstein's Eqs. (\ref{ob1}) and (\ref{ob2}) are valid then the terms in square brackets in
Eqs. (\ref{pr1ac}) and (\ref{pr2ac}) should be equal to constant $C$. This implies that the
following two equations
\begin{equation}
M_vI_v-M_sI_s =C,\label{fin1}
\end{equation}
\begin{equation}
(M_v-m_v)I_v'-(M_s-m_s)I_s'=C \label{fin2}
\end{equation}
should hold true. 
Now if the emission does not affect internal energy per unit mass of the body as viewed in different reference frames $S_s$ and $S_v$, respectively, then
\begin{equation}
I_s=I_s', \quad I_v=I_v'.
\end{equation}
Substituting it into Eq. (\ref{fin1}) and subtracting resulting equation from Eq. (\ref{fin2}) yield \textit{hidden condition}
\begin{equation}
m_sI_s'=m_vI_v'.
\end{equation}
This condition suggests that values for internal energy associated with decrease in mass of the body as measured in $S_s$ and $S_v$, respectively, are identical. \textit{Consequently the hidden condition means that internal energy of any body should have identical value when measured in $S_s$ and $S_v$ and which is perfectly valid within the framework of relativity}. In view of this, then we have
\begin{equation}
M_sI_s=M_vI_v, \quad M_sI_s'=M_vI_v'
\end{equation}
and substituting it in Eqs. (\ref{fin1}) and (\ref{fin2}) we obtain
\begin{equation}
C=0.
\end{equation}
The obtained value of $C=0$ suggests that Einstein could have avoided invoking argument ($EA$) for writing Eqs. (\ref{ob1}) and (\ref{ob2}) and using $C$ altogether and still could have derived the mass-energy equivalence relation by invoking the above mentioned hidden condition related to internal energy. This means that the new derivation should obtain Eq. (\ref{e3}) from Eq. (\ref{eq2a}) without using Eqs. (\ref{ob1}) and (\ref{ob2}) and which is now presented below. 

Now consider Eqs. (\ref{pr1ac}) and (\ref{pr2ac}) which were obtained by writing total energy of body $E$ and $H$ as summation of kinetic and internal energies. The obtained hidden condition as described above is a valid condition within the framework of relativity. So using this condition in Eqs. (\ref{pr1ac}) and (\ref{pr2ac}), we obtain \textit{exact} equations
\begin{equation}
H_0-E_0=K_0, \label{vik01}
\end{equation}
\begin{equation}
H_1-E_1=K_1 \label{vik02}
\end{equation}
which do not include any $C$ and are derived without invoking EA. 
These Eqs. (\ref{vik01}) and (\ref{vik02}) further yield
\begin{equation}
(H_0-E_0)-(H_1-E_1)=K_0-K_1 \label{my01}
\end{equation}
without invoking $EA$ and Einstein's Eqs. (\ref{ob1}) and (\ref{ob2}), thus making Ives's criticism a void. Then, Eqs. (\ref{my01}) and (\ref{eq2a}) yield
\begin{equation}
K_0-K_1=L[\frac{1}{\sqrt{1-v^2/c^2}}-1]
\end{equation}
which is identical to Eq. (\ref{e3}) and from which mass-energy equivalence relation Eq. (\ref{e4}) follows.


\end{document}